\documentclass[sn-mathphys]{sn-jnl}

\usepackage{bm}
\usepackage{mathtools}

\jyear{2021}%

\theoremstyle{thmstyleone}%
\theoremstyle{thmstyletwo}%

\theoremstyle{thmstylethree}%

\begin{document}

\title[Machine learning exchange-correlation functionals]{Development of exchange-correlation functionals assisted by machine learning}


\author*[1,2]{\fnm{Ryo} \sur{Nagai}}\email{ngrttt@gmail.com}

\author*[3]{\fnm{Ryosuke} \sur{Akashi}}\email{ryosuke.akashi31@gmail.com}

\affil*[1]{\orgdiv{Department of Physics, Japan}, \orgname{the University of Tokyo}, \orgaddress{\street{7-3-1 Hongo}, \city{Bunkyo-ku}, \postcode{1130033}, \state{Tokyo}, \country{Japan}}}

\affil[2]{\orgdiv{Institute for Solid State Physics}, \orgname{the University of Tokyo}, \orgaddress{\street{1-1-1}, \city{Kashiwa}, \postcode{2778581}, \state{Chiba}, \country{Japan}}}

\affil[3]{\orgdiv{Quantum Materials and Applications Research Center}, \orgname{National Institutes for Quantum Science and Technology}, \orgaddress{\street{2-10}, \city{Oookayama, Meguro-ku}, \postcode{1520033}, \state{Tokyo}, \country{Japan}}}


\abstract{With the recent rapid progress in the machine-learning (ML), there have emerged a new approach using the ML methods to the exchange-correlation functional of density functional theory. In this chapter, we review how the ML tools are used for this and the performances achieved recently. It is revealed that the ML, not being opposed to the analytical methods, complements the human intuition and advance the development toward the first-principles calculation with desired accuracy.}




\maketitle

\section{Introduction}\label{sec1}

A holy grail in the materials science is the general theoretical method to emulate chemical reactions. For this purpose we need to evaluate the energy of electrons distributed over the ionic configuration, with accuracy of the reaction, typically of order of kcal/mol. We already know what equation to solve--the Schr\"odinger equation 
\begin{align}
H=\sum_{n}\left[-\frac{\nabla^2}{2}+V_{\rm ion}({\bm r}_{n})\right]+\frac{1}{2}\sum_{m\neq n}\frac{1}{\|{\bm r}_{m}-{\bm r}_{n}\|},
\\
H\Psi({\bm r}_{1},{\bm r}_{2},\cdots)=E\Psi({\bm r}_{1},{\bm r}_{2},\cdots).
\label{eq:Schroedinger}
\end{align}
Here, $V_{\rm ion}$ denotes the one-body potential due to the distribution of ions. Arguably this is the theory of everything of electrons (apart from the relativistic, ion vibrational and external perturbation effects). But this equation is too difficult to solve exactly for realistic many-electron systems because its computational cost is of order of $\exp(N)$, with $N$ representing the system size. An alternative but yet exact route to the first-principles electronic structure calculation is opened by the density functional theory~\cite{Hohenberg-Kohn,Kohn-Sham}. This theory allows one to recast the original many-body problem to a one-body self-consistent equation, Kohn-Sham (KS) equation, as introduced later. 

The self-consistent potential entering the KS equation, exchange-correlation potential $V_{\rm xc}$ is exactly defined, but in a way that requires computational cost equivalent to the exact solution of the Schr\"odinger equation. It therefore has to be approximated for practical use. The systematic way to improve the approximate formula, however, has been lacking. Construction of the approximate forms of $V_{\rm xc}$ has been executed by assembling exactly computable asymptotic behaviors of the electronic systems, but, with the lack of the systematic derivation, finally we have to rely on a human intuition and empiricism to decide how to integrate them into a unified formula. 

Recently, a new trend has emerged in the development of $V_{\rm xc}$: Applying the machine learning (ML) to $V_{\rm xc}$. The ML methods enable us to implement a numerically calculable function that imitates some other function that is difficult to run on the computer. Typically the ML targets human functions like pattern recognition, while it can also be utilized to make a computational shortcut, by building an approximate model that accurately reproduces any other numerically demanding formula, algorithm, procedure, etc. An idea thus arose, that we may construct calculable models of $V_{\rm xc}$ by machine-learning numerous specific calculable cases. This may seem to be an extremely empirical approach that opposes to the standard theoretical one based on idealization. In reality, the ML has been found to be capable of complementing the human, by providing tools to explore models that well represent nonideal systems where analytical approach does not apply, and accelerating the development of approximate $V_{\rm xc}$.

In this book chapter, we review a recent studies of formulating approximate forms of $E_{\rm xc}$ and $V_{\rm xc}$ using the ML. In Sec.~\ref{sec:DFT}, we append a pedagogical review of the foundation of DFT to clarify what to machine-learn. In Sec. ~\ref{sec:MLtoXC} we introduce the basic concept of applying ML to $V_{\rm xc}$, explain the approaches, and make a survey of the accuracy achieved by the ML approach. Section ~\ref{sec:summary} is devoted to the summary and future perspectives.

\section{Basics of density functional theory}
\label{sec:DFT}
\subsection{Hohenberg-Kohn theorem}
The heart of the density functional theory is the concept that the charge density in the ground state,
\begin{eqnarray}
n({\bm r})
=
\int d{\bm r}_{2} d{\bm r}_{3}\cdots \| \Psi_{\rm GS}({\bm r}, {\bm r}_{2}, {\bm r}_{3}, \cdots)\|^2
,
\end{eqnarray}
determines all the physical properties of the quantum system. This has been formalized as the two Hohenberg--Kohn theorems~\cite{Hohenberg-Kohn}. According to the Schr\"odinger equation Eq.~(\ref{eq:Schroedinger}), when the ionic potential distribution $V_{\rm ion}$ is given, the ground state wavefunction $\Psi$ and charge density $n$ is unambiguously determined. Hohenberg and Kohn have proved the converse--when $n$ is given, $V_{\rm ion}$ that reproduces the given $n$ in the ground state is unambiguously determined (up to arbitrary constant). Consequently $\Psi_{\rm GS}$, and everything defined with $\Psi_{\rm GS}$, are determined via the Schr\"odinger equation. The related quantity $f$ is thus regarded as a functional of $n$, $f[n]$ (Hereafter we express the dependence as a functional of $n$ as $[n]$). The other theorem is the variational principle. Define the functional $F[n]$ as
\begin{eqnarray}
F[n]
=
\langle \Psi_{\rm GS}[n]\|\left\{\sum_{n}\left(-\frac{\nabla^2_{n}}{2}\right)+\sum_{n\neq m}\frac{1}{2\|{\bm r}_{n}-{\bm r}_{m}\|}\right\}\|\Psi_{\rm GS}[n]\rangle
\label{eq:univ-F}
\end{eqnarray}
When an external potential $V_{\rm ion}$ is given, the following inequality holds
\begin{eqnarray}
E[n]\equiv
F[n]
+\int d{\bm r} n({\bm r})V_{\rm ion}({\bm r})
\geq
E[n_{0}],
\label{eq:variational}
\end{eqnarray}
where $n_{0}$ satisfies $V_{\rm ion}=V_{\rm ion}[n_{0}]$. The equality holds only when $n=n_{0}$. This theorem yields that the ground state charge density at a given ionic potential can be obtained by solving the variational problem. With these theorems, the solution of the Schr\"odinger equation for the ground state can be recast to optimization of a scalar function on ${\bf R}^3$ space, $n$. The latter procedure can be executed with a surprisingly cheap computational cost, typically of order $O(N^3)$, if an explicit form of $F$ as a functional of $n$ is available. In this procedure, we do not have to explicitly treat the wave character of the electron. Although this approach has been pursued as orbital-free- (OF-) DFT, a more practical method has become the standard, which incorporates the concept of orbital for convenience as detailed below.

\subsection{Kohn-Sham equation}
The roadblock for the practical OF-DFT is that the kinetic energy, which largely reflects the wave character of electrons and is difficult to represent as a functional of $n$, accounts for a large fraction of the total energy. While the development of the functional form of $E_{\rm kin}$ has long been conducted, an alternative approach has been initiated by Kohn and Sham~\cite{Kohn-Sham}. In their theory, an auxiliary ``non-interacting" system, reproducing $n$ in the interacting ground state, is formulated
\begin{eqnarray}
\left[
-\frac{\nabla^2}{2}
+
V_{\rm eff}[n]({\bm r})
\right]
\varphi_{i}({\bm r})
=\varepsilon_{i}\varphi_{i}({\bm r}),
\label{eq:KS-eq}
\end{eqnarray}
where the charge density is given by
\begin{eqnarray}
n({\bm r})
=
\sum_{i}\theta(\mu-\varepsilon_{i})\|\varphi_{i}({\bm r})\|^2
\equiv
\sum_{i:{\rm occ.}}\|\varphi_{i}({\bm r})\|^2
.
\end{eqnarray}
Hereafter summation $\sum_{i:{\rm occ.}}$ runs over the occupied states, whose eigenvalues are smaller than a chemical potential $\mu$. The effective potential has a form
\begin{eqnarray}
V_{\rm eff}[n]( {\bm r})
=V_{\rm ion}({\bm r})+\int d{\bm r}' \frac{n({\bm r}')}{\|{\bm r}-{\bm r}'\|}
+
V_{\rm xc}[n]({\bm r}).
\end{eqnarray} 
The second (Hartree) term represents the classical Coulomb potential effect and the third, the exchange-correlation potential, includes the remaining interaction effects. Equation (\ref{eq:KS-eq}), the Kohn-Sham equation has $n$ in the right hand side and has to be therefore solved self-consistently.

$V_{\rm xc}$ is defined by
\begin{eqnarray}
V_{\rm xc}[n]({\bm r})
=
\frac{\delta E_{\rm xc}[n]}{\delta n({\bm r})}
\label{eq:Vxc}
\end{eqnarray}
with $E_{\rm xc}[n]$ being the quantity named exchange-correlation energy. $E_{\rm xc}[n]$ is defined with the following equality
\begin{eqnarray}
E_{\rm GS}[n]
=
T_{\rm s}[n]
+\int d{\bm r}V_{\rm ion}[n]({\bm r})n({\bm r})
+E_{\rm H}[n]+E_{\rm xc}[n]
.
\label{eq:EGS-Exc}
\end{eqnarray}
$E_{\rm GS}[n]$ is the ground-state energy as a functional of $n$. $T_{\rm s}[n]$ is the ``kinetic energy" defined by
\begin{eqnarray}
T_{\rm s}[n]
={\rm min}_{\|\Psi\|^2=n}\langle \Psi \|\sum\left[-\frac{\nabla^2_{n}}{2}\right]\|\Psi\rangle,
\end{eqnarray}
where $\Psi$ is restricted to be the Slater determinant reproducing the given $n$ as its squared norm. $E_{\rm H}[n]$ is the Hartree energy
\begin{eqnarray}
E_{\rm H}[n]
=\frac{1}{2}\int d{\bm r}d{\bm r}'
\frac{n({\bm r})n({\bm r})}{\|{\bm r}-{\bm r}'\|}.
\end{eqnarray}
The exchange-correlation energy $E_{\rm xc}$ is defined as the residual part of the total energy, which is also a functional of $n$. Note that the interacting ground-state energy $E_{\rm GS}[n]$ can be rewritten as
\begin{eqnarray}
E_{\rm GS}[n]
=
\sum_{i:{\rm occ.}}\varepsilon_{i}
+
E_{\rm H}[n]+E_{\rm xc}[n]-\int d{\bm r} n({\bm r})V_{\rm xc}[n]({\bm r})
.
\label{eq:KS-EGS}
\end{eqnarray}
We can thus calculate the ground-state charge density and total energy using Eqs.~(\ref{eq:KS-eq}) and (\ref{eq:KS-EGS}), provided that a specific expression of $E_{\rm xc}$ is available.

\subsection{Approximating the exchange-correlation functional}
Once the exact formula of $E_{\rm xc}[n]$ is obtained, apparently we do no longer have to solve the Schr\"dinger equation for calculating the ground state charge density and total energy. However, such formula is currently unavailable. Because the definition of $E_{\rm xc}[n]$ involves $E_{\rm GS}[n]$, which is the exact ground state energy obtained through the Schr\"odinger equation, its analytical formula is probably impossible to write in an efficiently calculable form.

Practical use of the KS equation is realized by constructing calculable approximate forms of $E_{\rm xc}[n]$. The most useful approximation is the semi-local approximation~\cite{Hohenberg-Kohn}, in which we introduce the local exchange-correlation energy density $\varepsilon_{\rm xc}$ by
\begin{eqnarray}
E_{\rm xc}[n]
=\int d{\bm r}n({\bm r})\varepsilon_{\rm xc}[n]({\bm r}) 
\label{eq:int_exc}
\end{eqnarray}
and simplify that $\varepsilon_{\rm xc}[n]({\bm r})$ depends only on the charge density and its derivatives at ${\bm r}$. Within this approximation, $\varepsilon_{\rm xc}[n]({\bm r})\simeq \varepsilon_{\rm xc}(n({\bm r}))$ gives the local density approximation (LDA). This approximation becomes exact in the uniform electron gas limit as $n({\bm r})=n={\rm const.}$ Vosko, Wilk and Nusair~\cite{LDA-VWN} and Perdew and Zunger~\cite{LDA-PZ} referred to the accurate calculation of $\varepsilon_{\rm xc}$ in the homogeneous electron gas obtained by the diffusion Monte Carlo method~\cite{Ceperley-Alder} and fitted their model functions. The model functions VWN and PZ, after the authors' names, are used in practice today. The next order correction within the semilocal approximation is included by adding a variable $\nabla n({\bm r})$; this is called generalized gradient approximation (GGA). The GGA-PBE functional~\cite{GGA-PBE} is a famous model functional within this level. To construct this model functional, one refers to rigorous formulas that have to be satisfied by $\varepsilon_{\rm xc}$; asymptotic equalities and inequalities in some limits like the uniform dense limit, and scaling relations. The model function is constructed with a minimal number of parameters and those are determined so that the model complies with those formulas. 

Further improvement has been conducted by considering the higher-order derivatives of $n({\bm r})$ and explicitly including the KS orbitals. The modern meta-GGA~\cite{becke1998new, perdew1999accurate, TPSS, SCAN} and hybrid functionals~\cite{B3LYP, PBE0} have been proposed from such improvement. For example, the hybrid functionals incorporates by some fraction the exact exchange energy formulated by the KS orbitals
\begin{eqnarray}
E_{\rm x}^{\rm EXX}
=
-\frac{1}{2}   \sum_{ij}^{\rm occ.} \! 
\int  \! \!  \! \int   \! d{\bm r}d{\bm r}' \varphi^{\ast}_{i}({\bm r})\varphi^{\ast}_{j}({\bm r}')
\frac{1}{\|{\bm r} \! - \! {\bm r}'\|}\varphi_{i}({\bm r}')\varphi_{j}({\bm r})
.
\end{eqnarray}
See e.g.~\cite{Kummel-Kronik,Head-Gordon} for a review. Precisely speaking, the approximations of this class is beyond the original KS theory in that the corresponding exchange-correlation potential is non-local: Namely, the matrix element $\left[V_{\rm xc}[n]\right]_{{\bm r}, {\bm r}'}$ is not proportional to $\delta({\bm r}, {\bm r}')$. Such extension is in fact justifiable as the generalized kohn-Sham theory~\cite{Generalized-KS}.

The goal of the development of the DFT (for the condensed matter) is a unified computational scheme that achieves accuracy at the energy scale of chemical reactions. The target accuracy of scale $\sim 1{\rm kcal}/{\rm mol}$ is called chemical accuracy. The levels of approximations for the DFT toward this goal has been classified as steps in ``Jacob's ladder~\cite{Jacob-ladder}", where the respective steps from below correspond to the LDA, GGA, metaGGA, hybrid, and those depending on the unoccupied KS orbitals.

\subsection{Exact exchange-correlation functional}
The celebrated statement that the exact xc functional has been elusive needs some disclaimer. Actually we already have both well-defined formulas of $E_{\rm xc}$ and $V_{\rm xc}$ as functionals of $n$ and numerical algorithm to evaluate them. The problem is that we do not have a method that is computationally cheaper than directly solving the many-body Schr\"odinger equation. Here we review some numerically executable exact formalisms for the functionals, which would be useful, and are already in use somehow, for developing the ML functionals.

$V_{\rm xc}$, as a functional of a given charge distribution $n$, is determined by Eq.({\ref{eq:EGS-Exc}}) and derivative Eq. ({\ref{eq:Vxc}}). On the other hand, the one-to-one correspondence between $V_{\rm ion}[n]$ and $n$ allows us to numerically calculate $V_{\rm xc}$ by solving inversion problems. Here we remember that in the non-interacting system [Eq.~(\ref{eq:KS-eq})] the one-to-one correspondence holds between $V_{\rm eff}[n]$ and $n$ and in DFT $V_{\rm ion}[n]$ is not an input but a quantity derived from the given $n$. Wagner and coauthors~\cite{Wagner_ExactKS} have demonstrated the numerical algorithm strictly conforming to this input-output relation. They solved the inverse problems of the KS [Eq.~({\ref{eq:KS-eq})] and many-body Schr\"odinger equations [Eq.(\ref{eq:Schroedinger})] to determine $V_{\rm eff}[n]$ and $V_{\rm ion}[n]$, respectively. Subtracting the latter from the former yields $V_{\rm H}[n]+V_{\rm xc}[n]$. This algorithm, involving the solution of the Schr\"odinger equation, is not computationally advantageous but was useful for proving the converging nature of the exact KS self-consistent equations~\cite{Guaranteed-convergence-KS}.

If the ionic positions and exactly calculated $n$ are available, we can derive $V_{\rm xc}$ by only the inversion of the KS equation thanks to the one-to-one correspondence. This procedure is called KS inversion~\cite{Zhao-Inverse-KS}. The calculation cost of the KS inversion is in principle the order of the forward solution of the KS equation times number of iteration, but its numerical convergence has remained a cumbersome problem, especially in three dimensions~\cite{Kanungo_KSinv}.

An analytical expression is also available for $E_{\rm xc}$, that is called analytically continued fluctuation dissipation formula~\cite{Langreth_Perdew}
\begin{eqnarray}
E_{\rm xc}
&=&
E_{\rm x}^{\rm EXX}
\nonumber \\
&&-
\frac{1}{2}\int_{0}^{1}d\lambda
\int\frac{d\omega}{2\pi}
\int d{\bm r}d{\bm r}'
\frac{\chi_{\lambda}[n]({\bm r}, {\bm r}',i\omega)-\chi_{0}[n]({\bm r}, {\bm r}', i\omega)}{\|{\bm r}-{\bm r}'\|}
.
\end{eqnarray}
 Here, parameter $\lambda$ defines the system
\begin{eqnarray}
H_{\lambda}[n]
=
\sum_{n}\left[-\frac{\nabla^2_{n}}{2}+V_{\rm ion}({\bm r}_{n})
+V_{\lambda}[n]({\bm r}_{n})\right]
+\frac{\lambda}{2}\sum_{m\neq n}\frac{1}{\|{\bm r}_{m}-{\bm r}_{n}\|}
.
\end{eqnarray}
and $V_{\lambda}$ is defined to be the potential that reproduces the ground state density $n$, identical to that of $H=H_{\lambda=1}$ and therefore $V_{\lambda=1}=0$ and $V_{\lambda=0}=V_{\rm KS}$. $\chi_{\lambda}$ denotes the density-density response function calculated for the ground state of $H_{\lambda}$. The $\lambda$ integral is taken with the variable $n$ fixed. Here we notice that, for evaluating $\chi_{\lambda}$, we need to solve the interacting Schrodinger equation for various $\lambda$. An approximation has been derived by, for example, approximating $\chi_{\lambda}$ to the independent-particle polarization expressed by the KS orbitals. The random phase approximation, which belongs to the fifth step of the Jacob ladder, can describe the dispersion interactions like the van der Waals one, which propagate between the intermediate excited states.


\section{Development of ML-based XC functionals}
\label{sec:MLtoXC}
\subsection{ML to DFT: Concepts and pioneers}

From the theoretical structure of DFT seen above, we read its significant features. (i) Once the exact form of functional, $F[n]$, $E_{\rm xc}[n]$ or $V_{\rm xc}[n]( {\bm r})$, is formulated, we can calculate the ground state charge density and energy apparently without using the Schr\"odinger equation. (ii) The functionals are well defined as functionals of the charge density $n$ through rigid procedures, but not as an explicit analytical functional of $n$. (iii) The procedures involve a step of solving the Schr\"odinger equation, because of which it is almost impossible to faithfully execute for general many-electron systems. 

Here we notice that the DFT formalism imposes a crucial limitation to the functionals as input-output relations; vector in, vector out. Any specific distribution of $n$ can be numerically described as a vector (for example, $(n({\bm r}_{1}, n({\bm r}_{2}),\cdots)$ with arbitrary discretization in real space). Letting a specific vector $n$ through the functional yields a definite scalar or vector. The previously proposed functional forms, referring to the numerically exact uniform electron gas case and several asymptotes, are regarded to be approximations that converge to the exact functionals in  limited spaces formed by the $n$ vector. In their constructions, model functions are prepared and the few parameters are determined so that the asymptotes are satisfied.

From the above view, one can conceive an approach that is a straightforward but extreme extension of the standard ones; prepare an extremely flexible model function and determine numerous parameters so that the function becomes exact in maximally large region in the $n$ space. Here the machine learning (ML) provides us useful models which can implement any kinds of vector-to-vector functions. The neural network (NN), the ML model we focus on in this chapter, is defined with a repetitive linear and non-linear operations
\begin{eqnarray}
{\bm x}
\rightarrow
\cdots \phi(W_{2}\phi(W_{1}{\bm x}+{\bm b}_{1})+{\bm b}_{2}) \cdots
.
\end{eqnarray}
Here, $W_{1}, W_{2}, \cdots$ and ${\bm b}_{1}, {\bm b}_{2}, \cdots$ are parameter matrices and vectors, respectively, having arbitrary intermediate dimensions. Non-linear transformation $\phi$ acts on each vector component. It has been proved that, by increasing the intermediate dimensions, the NN can reproduce any continuous functions up to arbitrary accuracy; this is known as universal approximation theorem~\cite{Hornik-UnivApprox,Cybenko-UnivApprox}. This theorem ensures that the NN, though not being the exact functional, can approximate the exact one up to arbitrary accuracy.

The idea of approximating the exact functional $V_{\rm xc}[n]$ with a NN has been first exemplified by Tozer, Ingamells and Handy~\cite{Tozer-NN} in 1996. They calculated $n$ for some atoms with an accurate wave function method, derived $V_{\rm xc}$ by the KS inversion method, and optimized a few-parameter NN  as a mapping from $n({\bm r})$ to $V_{\rm xc}({\bm r})$ so that it well reproduces the example $n-V_{\rm xc}$ pairs (training). Their procedure can be scaled up straightforwardly, to the modern machine-learning approaches to the DFT functional.

As a pioneering work, a specific mention also proceeds to a work by Zheng and coauthors in 2004~\cite{Zheng2004}. They proposed to make the parameters in a hybrid functional, B3LYP~\cite{B3LYP1, B3LYP2}, system-dependent. They implemented the mapping from the atomic species and positions to the B3LYP parameters by the NN and trained it referring to experimentally observed atomization energy data of molecules.

The surge of the ML application to DFT in the recent decade seems to have been prompted by a work by Snyder and coauthors~\cite{snyder2012finding}. In a one-dimensional model system, they implemented the kinetic energy functional, $F[n]$ [Eq.~(\ref{eq:univ-F})] in the non-interacting limit, by the gaussian kernel regression, for possible execution of the orbital-free DFT. This model form does not need the model parameter optimization but needs to store the $n$-$F[n]$ pairs with various system parameters for application to unstored system parameter regimes. In this and their subsequent works~\cite{snyder2013orbital} the gaussian kernel regression model were used, though the NN is more frequently used in later studies.

The basic concepts of more recent ML applications to the DFT problems have already been put forward in those predecessors. Below we summarize the various ML applications to the task of calculating the electronic states on the basis of DFT.



\subsection{Levels of ML applications}
The electronic properties are determined {\it ab initio} by atomic positions through the Schr\"odinger equation. The ML scheme can in principle be applied to every steps in this procedure. Firstly, the whole procedure itself is well defined as a mapping from atomic positions ($+$ number of electrons) to electronic properties. One may therefore apply the ML to skip the solution of the Schr\"odinger equation by modeling the atomic position-to-energy mapping. Methods of this kind have been extensively explored for efficient molecular dynamics with the electronic effects considered (e.g., neural network potential in molecular dynamics~\cite{Behler-Parrinello}). Although such ML model, if efficiently trained, enables us to calculate the electronic properties with the cheapest computational cost, the model would be less transferable since it becomes atomic-species-dependent. We do not review the applications of this kind in detail in this chapter.

The fundamental mappings in DFT are independent of atomic kinds--the charge density $n$ does not regard on which atomic positions it distributes. Brockherde and coauthors~\cite{Burke-bypassing} implemented ML models for the mapping from $V_{\rm ion}$ to $n$, whose one-to-one correspondence has been proved as the Hohenberg-Kohn theorem~\cite{Hohenberg-Kohn}. Moreno, Carleo, and Georges~\cite{Georges_HKmap} and Denner, Fischer and Neupert~\cite{Denner_1d_correlation} attempted implementation of some of such functionals; $n$ to ground-state wave function and two-particle correlation function.

Successful implementation of the direct mapping from $n$ to universal functional $F[n]$ paves the way to the orbital-free (OF) DFT. This is one of the most active field for the ML application from the beginning of this decade~\cite{snyder2012finding, snyder2013orbital, Burke-bypassing,Seino_OFDFT2018,Imoto_OFDFT2021}. In OF-DFT, the ground state energy is calculated by directly minimizing Eq.~(\ref{eq:variational}) with respect to $n$ without solving the KS equation. The main issue is to approximate the kinetic energy as a functional of $n$, which constitutes the largest fraction of the total energy. The main subject in OF-DFT is accurate represention of the kinetic energy as a functional of $n$. 

Another approach respects accuracy than calculation speed, retaining the KS framework and model some steps from the KS equation to ground state charge density. At the expense of the computational cost of solving the KS equation, this approach takes advantage of treating the kinetic energy exactly as an operator on the (KS) wavefunction. This gives us two major rewards. First, the largest part of the total energy is treated exactly, on top of which the ML model can learn precise $n$ dependence of the remainder. Second, the kinetic energy operator suppresses spurious oscillation in the resulting $n$, significantly enhancing the transferability of the model; a sort of ``regulator~~\cite{NNVHxc}" of the machine. A typical ML application in this level is to the mapping from $n$ to $E_{\rm xc}$ or $V_{\rm xc}\equiv\delta E_{\rm xc}/\delta n$. Using $V_{\rm xc}[n]$ trained in some way, we solve the KS equation in application to untrained systems, as usual in the standard practice of the KS-DFT. The current chapter mainly concerns the applications at this level. 

\subsection{How to train the ML models}
There is a common strategy for training whatever functionals. We first prepare ``accurate" examples of input $\{n^{(i)}\}$ ($i$: index of training data)  and output value(s) of functional(s) for specific systems, $\{f^{(i)}\equiv f[n^{(i)}]\}$. We define a loss function by the difference between $\{f^{(i)}\}$ and the values of the same quantity approximately evaluated by the use of ML, $\{f^{(i)}_{\rm ML}\}$, and optimize the ML model so that the loss function is minimized. The method for generating the ``accurate" data should be numerically exact, as the approximate solvers for the training data limit the accuracy of the ML model.

The methods for the training data, like wave function theory-based methods, usually do not have the concept of exchange-correlation energy or potential. The training data of exchange-correlation potential as a functional of $n$ can be calculated by the Kohn-Sham inversion technique. Notably, the formula of $V_{\rm xc}[n]$ is enough for evaluating the total energy of the system by the KS equation, despite the total energy formula Eq.(\ref{eq:KS-EGS}) seemingly needs $E_{\rm xc}[n]$, not only $V_{\rm xc}[n]$. This can be made possible by tuning the constant term of $V_{\rm xc}[n^{(i)}]$, at the end of the inversion, so that the simple sum of the KS eigenvalues agrees to the total energy calculated by the accurate solver, $E^{(i)}$:
\begin{eqnarray}
E^{(i)}_{\rm GS}
=\sum_{j}\varepsilon_{j}^{(i)}
.
\end{eqnarray}
This is a practical usage of the Levy-Zahariev formalism~\cite{Levy_Zahariev2014}, where we introduce a scalar functional $c[n]$ in Eq.~(\ref{eq:KS-EGS}) as
\begin{eqnarray}
E_{\rm GS}
=
\sum_{i:{\rm occ.}}(\varepsilon_{i}+c[n])
+E_{\rm H}+E_{\rm xc}-\int d{\bm r}n({\bm r})\left[V_{\rm xc}[n]({\bm r})+c[n]\right]
\end{eqnarray}
and define $c[n]$ so that the latter terms vanish.

Although learning the $n$ to $V_{\rm xc}$ mapping as explained above is the straightforward way to apply ML to the KS equation, there are variations in the way  to model the functionals. Here we classify the modeling methods used in the recent studies, though these modelings are often used in combination.

\subsubsection{Learning $n \rightarrow V_{\rm xc}$ mapping}

Most generally, $V_{\rm xc}[n]$ is a function relating the entire distribution of $n$ to the entire potential. Nagai and coauthors demonstrated the training of $V_{\rm xc}[n]$ in a one dimensional two spinless electron system~\cite{NNVHxc}. The fully nonlocal form is little transferable since it does not apply to the cases with different space grids (range, spacing, etc). Moreover, such form requires large amount of training data because a single set of model parameters gives us only a single set of the training pairs.

These two problems are solved by limiting the model form so that the potential value depends only on the near distribution of $n$. First, a single global distribution of $n$ has various local $n({\bm r})$ values, which enables us to train the model with smaller amount of training data. Second, such a local model is applicable to systems with different sizes. Schmidt and coauthors implemented a model form of the exchange-correlation energy density $\varepsilon_{\rm xc}[n]({\bm r})$ that depend on density within a near region of ${\bm r}$ for a 1D model~\cite{Schmidt_1D_2019}. Lyabov, Akhatov and Zhilyaev showed that a similar near-region model can learn nonuniform systems to reproduce the traditional LDA-PZ and GGA-PBE functionals in three dimensions~\cite{Lyabov2020}. 

\subsubsection{Learning $n\rightarrow E$ mapping} 
Another direct way to train the functional is to use data of density and physical/chemical quantities such as atomization energy or ionization potential. In this strategy, $E_{\rm xc}$ rather than $V_{\rm xc}$ is conveniently modeled as a functional of $n$ (and $\varphi[n]$), using the semilocal form. The physical/chemical quantities are prepared by any database of experimental values or highly-accurate wavefunction theory such as coupled-cluster theory \cite{CCSD}. The loss function is  then defined with the difference of the reference quantity $\|Q^{(i)}-Q^{\rm ML}[n^{(i)}]\|$ for reference pairs $(n^{i}, Q^{(i)})$. The training dataset must include a large number of materials to train the ML parameters adequately, since the reference physical quantities are integrated values, which can only be obtained once per material. 

In the modelling of this type, the reference $n$ and energies are often generated with different approximations for saving of time. This would in principle suffer from a subtle error, though the performance is generally insensitive to the functional used for generating $n$. This could be thanks to an empirical fact that, with the KS calculation, the error due to inaccurate density tends to be smaller than that due to inaccurate functional form~\cite{dick_natcomm2020} ( density-driven and functional-driven errors in terms of Kim~\cite{Kim-Burke-errors2013}, respectively).  

Dick and Fernandez-Serra~\cite{dick_natcomm2020} built a model of Behler-Parrinello form~\cite{Behler-Parrinello} to correct the difference between energies calculated with a computationally cheap method (``baseline") and accurate one: The charge density is translated to atomic descriptors and each energy correction per atom is modeled as a function of the descriptors. The reference charge densities were generated with the KS calculation with the GGA-PBE~\cite{GGA-PBE} functional, whereas the accurate reference energy was calculated using the CCSD(T) method.

The absolute total energy is not preferable as a reference quantity since it shows large error; for example, the difference in total energies between those with PBE and full configuration-interaction method is a few hartrees. Reaction energies, defined as differences in total energies of the systems with different bonding forms, mostly show far smaller errors and are more useful for training. Kirkpatrick {\it et al.} \cite{Deepmind2021pushing} conducted the training of ML XC functional along the above strategy. They modeled $\Delta E_{\rm xc}$, the difference in $E_{\rm xc}$ between the reactant and product as a semilocal functional of $n$ and local exchange energy, and trained it with 1161 chemical reactions. They collected the reference pairs $(n^{(i)},\Delta E_{\rm xc}^{(i)})$ using the KS equation with the B3LYP~\cite{B3LYP} functional for $n$ and CCSD(T) for $\Delta E_{\rm xc}$, respectively.


\subsubsection{\label{sec:differentiable} Implicit training using KS equation}
One can also define the loss function by the difference between the accurately calculated $n$ and $n$ derived through the self-consistent KS equation with the ML exchange-correlation potential included. The model parameters are optimized so that the two densities agree. This approach is actually equivalent to that referring to $V_{\rm xc}$ because, to the HK theorem, $V_{\rm xc}$ reproducing the exact $n$ is exact and unique. A primary drawback of this method was that the gradient of the loss function with respect to the model parameters was unavailable. Nagai and coauthors~\cite{nagai2020completing} defined the loss function by the differences in $n$ and reaction energies and used a non-gradient stochastic method for updating the model parameters, which took large computational cost for the optimization. This problem has been solved by implementing the KS equation in differential programming. This idea has been first demonstrated in a 1D model system by Li and coauthors~\cite{li2021regularizer}. Kasim and Vinko~\cite{kasim2021learning} and Dick and Fernandez-Serra~\cite{dick2021highly} implemented a single NN for the molecules in 3D, where the derivatives iteratively back-propagates through the KS cycle. The differential programming, however, requires a large amount of memory, which makes it difficult to train on large dataset. 

\subsection{Imposing physical constraints\label{sec:physical_con}}
 Various analytic formulas obeyed by the exact $E_{\rm xc}$ have been revealed\cite{ uniform_density_scaling, spin_scaling, perdew2014gedanken, perdew1992accurate, levy1991density}; asymptotic expansion forms, inequalities, scaling relations, and so on. The uniform electron gas limit~\cite{Slater, LDA-VWN} is in particular important for the functionals to be accurate in solid systems~\cite{Perdew_prescription}. Ideally the approximate functionals should comply with all such known formulas. Controlling analytical structure of ML models as such is, however, not straightforward because of their flexibility. In this section, we review methods to impose physical constraints on ML-based XC functionals, which have been found to result in better accuracy and transferability.


\subsubsection{Learning data satisfying physical constraints}
The most direct way to impose physical constraint is to include data satisfying the physical constraints in the training data set. Kirkpatrick {\it et al.} \cite{Deepmind2021pushing} and Gedeon {\it et al.} \cite{gedeon2021machine} included training data in systems with fractional number of electrons into the training dataset, to make the XC functional complies to the ideal linear dependence \cite{perdew1982density}; namely, the ground-state total energy of the $M$-electron system $E(M)$ is exactly linear for $M$ being non-integer. This linearity condition is important for accurately treating the charge-transfer properties such as ionization potential and bandgap~\cite{Mori-Sanchez_PRL2008,Mori-Sanchez_Sci2008}, but most existing XC functionals do not satisfy it. Kirkpatrick {\it et al.} and Gedeon {\it et al.} generated data of noninteger number of electrons by interpolating those of integer number of electrons. They trained ML-based functionals with some non-locality using those data, and their XC functional well reproduced the behavior of total energy of fractional electrons. 

\subsubsection{Designing ML architectures \label{sec:pcNN}}
Some physical constraints represented by the limiting values, asymptotes and inequalities can be equipped by the NN models analytically. An example is the spin-scaling relationship for the spin-dependent exchange energy~\cite{spinscaling}
\begin{eqnarray}
E_{\rm x}[n_{\uparrow}, n_{\downarrow}]
=\frac{1}{2}
\left[E_{\rm x}[2n_{\uparrow}]+E_{\rm x}[2n_{\downarrow}]\right]
,
\end{eqnarray}
which may be applied to whatever model exchange functionals.

Nagai {\it et al.} has proposed a way to analytically impose a certain class of asymptotic constraints \cite{nagai_pcNN}:
Assume that we want to impose an asymptotic constraints that the  output of the ML function $f({\bf x})$ should converge to $f_0$ at ${\bf x} \rightarrow {\bf x_0}$. 
This constraints can be imposed by a post-processing operation $\hat{\theta}_{\{{\bf x}_0,f_0\}}$ defined as follows: 
\begin{equation}
    \hat{\theta}_{\{{\bf x}_0,f_0\}}[f]({\bf x})=f({\bf x})-f({\bf x}_0)+f_0.
    \label{eq:single}
\end{equation}
Even if we have multiple asymptotic constraints, where $f$ converges to $f_0^{(1)}, f_0^{(2)}, ... f_0^{(N_c)}$ at ${\bf x}_0^{(1)}, {\bf x}_0^{(2)},...,{\bf x}_0^{(N_c)}$, we can impose them simultaneously by combining Eq.~( \ref{eq:single}) in a way like the Lagrange interpolation:
\begin{align}
    \hat{\Theta}_{i=1,...N_c}[f]({\bf x})
    =
    \sum_{i=1}^{N_c} 
    \prod_{j\neq i}
    \frac{l_{i,j}({\bf x}-{\bf x}_j)}{l_{i,j}({\bf x}_i-{\bf x}_j)} 
    \hat{\theta}_{\{x_0^{(i)},f_0^{(i)}\}}[f]({\bf x})
    \label{eq:multi}
\end{align}
 $l_{i,j} (i, j=1,2,...N_c)$ are function satisfying $l_{i,j}({\bf x})=0$ at ${\bf x}=0$ and $l_{i,j}({\bf x}) \rightarrow 1$ at $\|{\bf x}\| \rightarrow \infty$. They suppress oscillations coming from the high-dimensional polynomial in Eq. \ref{eq:multi}. 
 
Most of the physical constraints are asymptotic-type, and thus they constructed an ML-based XC functional satisfying most of known physical constraints for the meta-GGA approximatin.  In their results, imposing physical constraints improved not only accuracy but also computational stability. They used dataset consists of molecular systems for the training. The SCF calculation using the XC functional constructed without physical constraints did not converge for some solid systems (metals, semiconductors), while that with physical constraints converged in the same speed as the conventional analytical functionals. Imposing physical constraints in this way is expected to regularize functional forms where the ML model cannot be trained enough and improves transferability.

In conjunction with the linearity condition, the exact $E_{\rm xc}$ shows a discontinuity in the derivative of $E_{\rm xc}$ with respect to the total number of electrons $M$ at the integer $M$~\cite{perdew1982density}:
\begin{eqnarray}
\frac{\partial E_{\rm xc}}{\partial M}(M+0)
-
\frac{\partial E_{\rm xc}}{\partial M}(M-0)
\equiv \Delta_{\rm xc}\neq 0
\end{eqnarray}
Gedeon {\it et. al.}~\cite{gedeon2021machine} applied a function of form $\|\sin \pi M\|$ so that the derivative discontinuity $\Delta_{\rm xc}$ is efficiently represented. They used this in combination with the aforementioned training data design to reproduce the ideal behavior.

The constraints represented by inequalities can be well enforced by using bounded non-linear transformation. Dick and Fernandez-Serra~\cite{dick2021highly} employed a function of form $\frac{a}{1-(a-1)\exp(-x)}-1$ with $a$ being a constant to impose the Lieb-Oxford bound for $E_{\rm x}$~\cite{LOB}, which defines lower bounds of exchange-correlation energy. 

\subsubsection{Designing the loss function}
The ML model is optimized by minimizing the loss function defined by the difference between the values calculated by the ML model and those in the reference data. By adding extra penalty terms to that, we can loosely enforce the constraints to the model. Pokharel {\it et al.}~\cite{Pokharel_constrain_arxiv} introduced a penalty term of form $\propto {\rm ReLu}(f(x)-a)$, with ${\rm ReLu}(x)$ being the rectified linear unit ${\rm ReLu}(x)={\rm max}(0, x)$, so that the model approximately satisfies an inequality condition $f(x)\leq a$. Another interesting approach has been put forward by Cuerrier, Roy and Ernzerhof~\cite{Cuierrier_exhole}. They modeled the negative-definite exchange-correlation hole by neural networks, defined an ``entropy" with that and targeted it as the loss. Although there is no guarantee that their maximum entropy principle is physical, this method makes the model smooth and likely enhance the transferability.


\subsubsection{Compressing input parameter space with physical constraints to improve transferability}
Hollingsworth {\it et al.} studied the use of coordinate-scaling condition  to improve  transferability in one-dimentional system \cite{kieron_exact_condition}. The scaling condition relates the values of density functional between densities in different length scales:
\begin{align}
T_s[n_\gamma] = \gamma^2 T_s[n], \label{eq:Tconvert} \\
n_\gamma(r) = \gamma^3n(\gamma r), \label{eq:nconvert}
\end{align}
where $T_s$ represents the kinetic energy functional of non-interacting electron.
First, the ML-based kinetic energy functional is trained using dataset consisting of density in a specific scale. When they applied the trained model to data in a different scale, the scaling parameter was adapted to the value of the training data by Eq.\ref{eq:nconvert}. The output of ML $T_s[n]$ is then converted to that of original density scale by Eq. \ref{eq:Tconvert}. They showed that the prediction accuracy with this method was better than directly predicting $T_s[n]$ of density in different scales.


\subsection{Performance of ML-based XC functionals}
Here we quote some state-of-the-art results accomplished by the ML-based exchange-correlation functionals. Nagai {\it et al.} demonstrated the construction of NN-based meta-GGA functional using dataset including atomization energy and density of 3 molecules \cite{nagai2020completing}. Its mean absolute error (MAE) for atomization energy of 147 molecules was 4.7 kcal/mol, while that of conventional meta-GGA functional with best performance (M06-L \cite{M06L}) was 5.2 kca/mol. Fig. \ref{fig:nagai} exhibits the accuracy of energy-related quantities and density. Later, they improved the accuracy by incorporating physical constraints into ML model architecture (See Section \ref{sec:pcNN}): its MAE for the same 144 molecules was 3.6 kcal/mol \cite{nagai_pcNN}.
\begin{figure}[t]
\centering
\includegraphics[width=1.0\columnwidth]{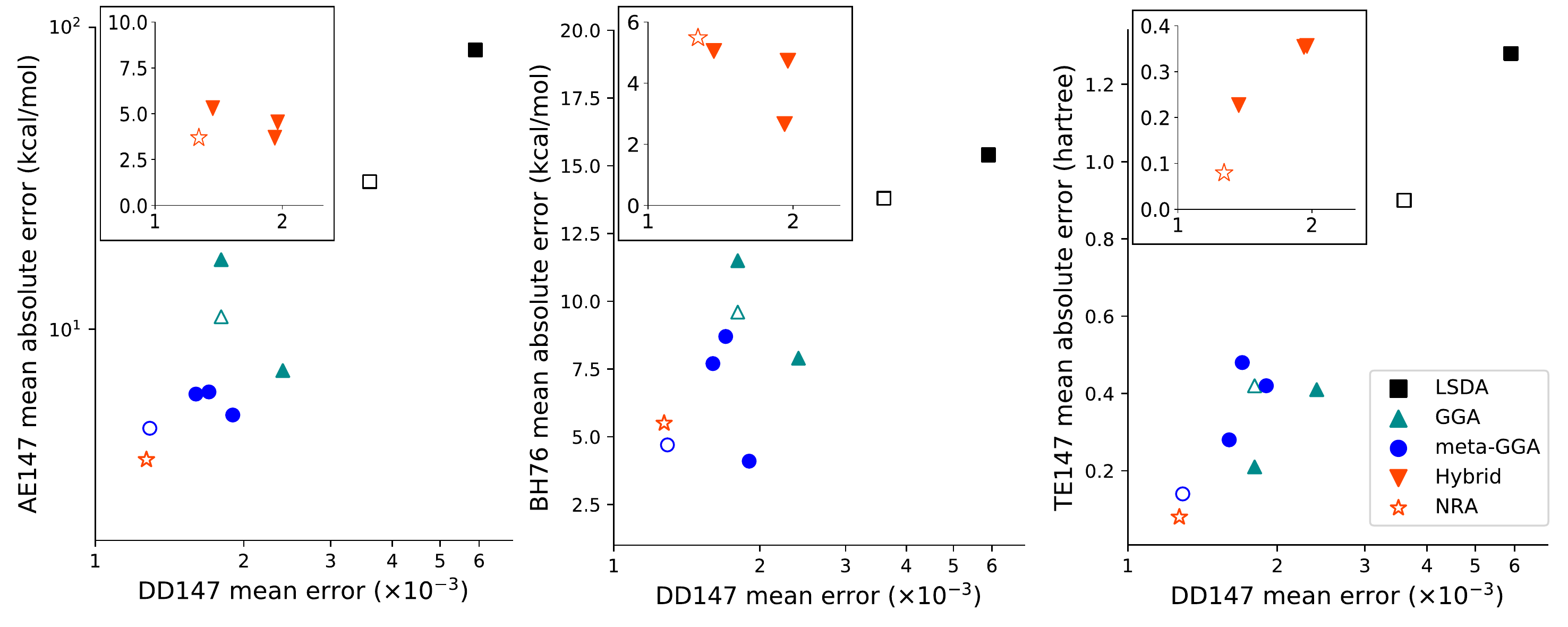} 
\caption{Relations between accuracy of density and that of atomization energy, density distribution, total energy, cited from Ref. \cite{nagai2020completing}. The panels represent mean absolute error (MAE) for atomization energy of 147 molecules (AE147), barrier height of 76 chemical reactions (BH76), and total energy  of 147 molecules (TE147) against MAE for density distribution of 147 molecules (DD147). The closed and open markers represent the accuracy of existing and the NN-based functionals, respectively. Legend “NRA” represents an improved approximation with a non-local density descriptor, constructed using NN.}
\label{fig:nagai}
\end{figure}

Li {\it et al.} implemented differentiable KS cycle and performed training with solving KS cycle (Section \ref{sec:differentiable}) for 1D molecules \cite{li2021regularizer}. Fig. \ref{fig:dens_and_pot} exhibits predicted effective potential $v_s$ ($=V_{\rm eff}$ in Eq.~(\ref{eq:KS-eq})) and the density $n$ obtained from solving KS equation on $v_s$. The results well reproduce the exact $v_s$ and $n$, and we can see that the XC functional trained using density and energy in the loss function (Fig. \ref{fig:dens_and_pot} (a)) had better accuracy than that using only energy (Fig. \ref{fig:dens_and_pot} (b)).
\begin{figure}[t]
\centering
\includegraphics[width=1.0\columnwidth]{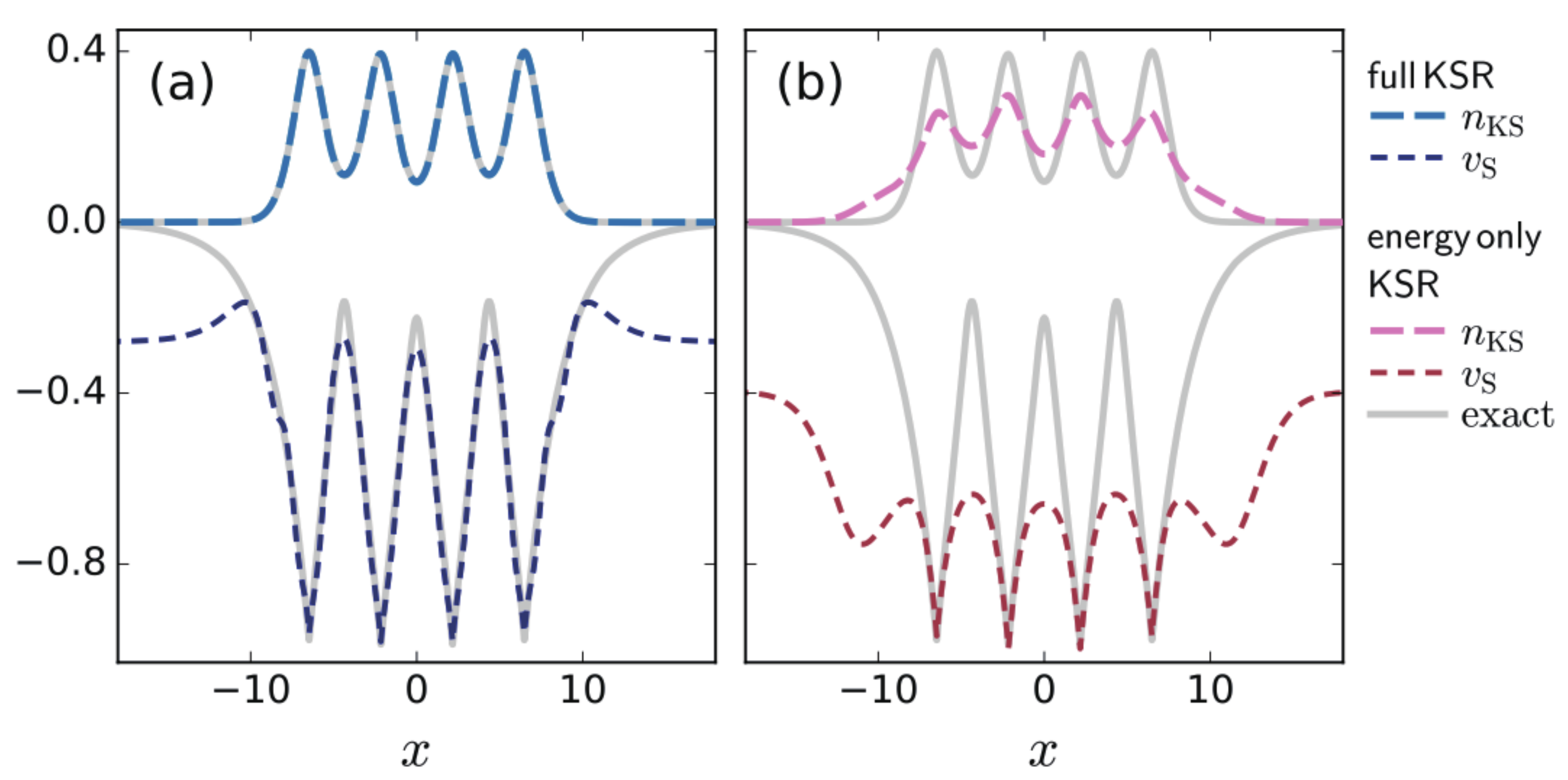} 
\caption{Density and KS potential of H4 from neural XC functionals trained with (a) energy and density loss and (b) energy loss only, cited from Ref. \cite{li2021regularizer}.}
\label{fig:dens_and_pot}
\end{figure}

Kasim {\it et al.} implemented the training using differentiable KS equation on 3D molecules \cite{kasim2021learning}. They used 4 molecules each for training and validation, and MAE for 104 molecules of their NN-based GGA functional  was 7.4 kcal/mol while that of GGA-PBE~\cite{GGA-PBE} was 16.5 kcal/mol. Dick {\it et al.} also demonstrated NN-based meta-GGA XC functional using differentiable KS equation \cite{dick2021highly}. They used AE, IP, barrier height (BH) and density of tens of molecular systems as the training dataset.
Table \ref{tab:Dick} compares their functional and standard conventional functionals in on the MAE for hundreds of molecular systems. 

\begin{table}[t]
\centering
 \caption{Mean absolute errors of atomization energies (AE) for the W4-11 dataset, barrier heights (BH) for BH76, decomposition energies (DE) for MB16-43, and densities ($\epsilon_{n}$) for G2/97, cited from Ref. \cite{dick2021highly, goerigk2017look}. “xc-diff” represents meta-GGA NN-based XC functional proposed by Ref. \cite{dick2021highly}. }
 \label{table:dick_pf}
  \begin{tabular}{lcccccc}
  \hline \hline
    & AE & BH & DE &WTMAD-2 & $\epsilon_{n}\times 10^3$\\
    & (kcal/mol) & (kcal/mol) & (kcal/mol) & (kcal/mol) & - \\
   \hline 
   RPBE   \cite{RPBE}& 8.3 & 9.0  & 50.8 & 10.5 & 8.8 \\
   B97    \cite{B97}& 4.7 & 7.3  & 36.1 & 8.6 & 7.0 \\ 
   OLYP   \cite{OLYP}& 9.9 & 8.5  & 29.0 & 8.5 &10.1\\ 
   revPBE \cite{revPBE}& 7.6 & 8.3  & 27.1 & 8.4 &9.4 \\
   M06L   \cite{M06L}& 4.4 & 3.9  & 63.3 & 8.6 &9.4 \\
   revTPSS\cite{revTPSS}& 5.7 & 8.9  & 36.7 & 8.4 &7.9 \\
   SCAN   \cite{SCAN}& 4.1 & 7.8  & 17.8 & 8.0 &6.2\\
   xc-diff& 3.5 & 6.5  & 22.7 & 7.3 &5.2 \\
   PBE0   \cite{PBE0}& 3.7 & 5.0  & 15.9 & 6.4 &5.7\\
   B3LYP  \cite{B3LYP}& 3.4 & 5.7  & 24.8 & 6.5 &8.3\\
   M05-2X \cite{M05-2X}& 4.0 & 1.7  & 26.3 & 4.6 &7.5 \\
   $\omega$B97X-V\cite{omegaB97X-V}& 2.8 & 1.8  & 32.5 & 4.1 &5.0 \\
   \hline \hline
  \end{tabular}
  \label{tab:Dick}
\end{table}

As mentioned in the previous chapter, Kirkpatrick {\it et al.} included fractional electron/spin systems in their training dataset. Furthermore, they adopted a local HF features, which are defined as range-separated exact exchange energy density
\begin{eqnarray}
e_{\omega \rm HF}^{\sigma}({\bm r})
=
-\frac{1}{2}   \sum_{ij}^{\rm occ.} \! \varphi^{\ast}_{i}({\bm r}) \varphi_{j}({\bm r}) \int   \! d{\bm r}' 
\frac{{\rm erf}(-\omega\|{\bm r} \! - \! {\bm r}'\|)}{\|{\bm r} \! - \! {\bm r}'\|}\varphi^{\ast}_{j}({\bm r}') \varphi_{i}({\bm r}')
,
\end{eqnarray}
with $\omega = 0.4$ and $\infty$.
Those features are expected to include information of non-local interaction. \cite{Deepmind2021pushing}. For their training, they used 1161 data including atomization, ionization, electron
affinity, and intermolecular binding energies
of small molecules, and 1074 data
represented fractional electron and spin systems of atoms. In one of the tests they performed, the MAE was 1.66 kcal/mol for the QM9 dataset \cite{QM9} containing 133,857 chemical reactions, while the MAE of an existing hybrid functional called $\omega$B97X was 2.13.

 A factor that contributes to the high transferability seen in those results is the use of local descriptor of electron density. When applying a constructed XC functional to a material, the value of XC energy density can be well predicted if the similar local density distribution is included in the training dataset, regardless of the kinds of ions. We therefore do not need to have exactly the same elements of the target material in the training dataset.

Another factor contributing to the transferability is a mathematical property of the KS equation. When an ML model is applied to local density distribution outside the region of the training dataset, there would be error in predicted XC energy density. As a result, there would occur spurious fluctuations in the XC potential. However, the kinetic energy term in the Kohn-Sham equation (Eq.~(\ref{eq:KS-eq})) has an effect of penalizing the non-smooth solution, and thus we can obtain smooth KS orbital (detailed discussions are in Ref. \cite{NNVHxc}).


\section{\label{sec:summary}Summary and future perspectives}
In this chapter, we introduced methods for constructing ML-based XC functionals for the Kohn-Sham equation. Machine-learning models such as NNs are trained with highly accurate electronic structure data of real materials to construct the XC functional. Although data-driven techniques of ML enable construction without physical constraints, the inclusion of physical constraints improves the accuracy and computational stability. There exists some choices in the way for the training. One can directly use the pair of electron density and energy as data, while incorporating the self-consistent cycle of the Kohn-Sham equation into the training procedure is also an effective way to ensure the numerical stability of constructed XC functional.
The trained machine learning model can be used by substituting it into KS equation for any materials. Since the electron density, which exists in common in any materials, is input of ML model, once an XC functional is well-trained, it would yield accurate prediction for various materials. In that sense, ML techniques is expected to be effective for the improvement of DFT.

In some studies ML-based XC functionals have outperformed the standard ones constructed by human. By using flexible ML models and using data of real materials for the  training, the ML-based XC functional is expected to learn the essential features of the electronic interaction effects which cannot be represented by existing physical theories. 



The goal of this field is to realize versatile XC functionals which is universally accurate for any materials. The ML models are far more flexible than the conventional analytical models, and this feature can be exploited for a highly transferable XC functional. However, at present, using data of small molecules is the mainstream, while learning with data of large materials has not been realized. It will be important to establish a method to utilize data of experiments as well as to expand the highly accurate electronic structure data set for various materials.

Meanwhile, specializing in a particular system would also be a practical application. For the difficult systems whose essential properties of their electronic states cannot be treated analytically, conventional construction relying on analytical way seems to fail. The ML-based construction is also expected to be effective for that cases because they can systematically extract the essence of electron interaction from dataset.

The construction of a functional by ML may be useful for understanding physics. For example, by limiting types of materials used in training and evaluating the performances of constructed ML-based XC functionals for different type of materials, the similarity in electronic structures of the two systems can be analyzed. It would also be possible to investigate which physical constraint contributes most to the accuracy by comparing differently constrained functionals in the way introduced in Section \ref{sec:physical_con}.

In this chapter, we introduced the construction of density functionals for ground state electron systems, but ML can be applied to other systems, such as time-dependent DFT \cite{suzuki2020machine} or classical DFT for solvent systems \cite{shang2019classical}. Through the accumulation of studies for those problems, we expect that a systematic way to complement unknown functionals in general variational problems would be established.





\bibliography{reference}


\end{document}